\documentclass[preprint,preprintnumbers,amsmath,amssymb]{revtex4-1}
\usepackage{txfonts}
\usepackage{amsfonts}
\usepackage{amsmath}
\usepackage{graphicx} 
\usepackage{dcolumn}  
\usepackage{bm}
\usepackage{overpic}
\usepackage{color}
\usepackage{multirow}
\newcommand{\beq}[1]{  \\ {\tiny ({#1})}   \begin{equation} \label{#1} }
\newcommand{\beqa}[1]{ \\ {\tiny ({#1})}   \begin{eqnarray} \label{#1} }
\newcommand{\eeq}{\end{equation}}
\newcommand{\eeqa}{\end{eqnarray}}

\begin{document}
\newcommand{\rf}[1]{(\ref{#1})}
\newcommand{\bfomega}{ \mbox{\boldmath{$\omega$}}}

\title{Stress-dependent normal mode frequencies from the effective
  mass of granular matter}


\date{\today}

\author{Yanqing Hu$^{1}$, David L. Johnson$^2$, John J. Valenza$^2$,
  Francisco Santibanez$^{1,3}$, Hern\'an A. Makse$^1$}

\affiliation{ $^1$ Levich Institute and Physics Department, City
  College of New York, New York, New York 10031, USA \\ $^2$
  Schlumberger-Doll Research, One Hampshire, Cambridge, Massachusetts
  02139, USA \\ $^3$ Departamento de F\'isica, Universidad de Santiago de Chile, Av. Ecuador 3493, Santiago, Chile.}

\begin{abstract}

A zero-temperature critical point has been invoked to control the
anomalous behavior of granular matter as it approaches jamming or
mechanical arrest. Criticality manifests itself in an anomalous
spectrum of low-frequency normal modes and scaling behavior near the
jamming transition. The critical point may explain the peculiar
mechanical properties of dissimilar systems such as glasses and
granular materials.  Here, we study the critical scenario via an
experimental measurement of the normal modes frequencies of granular
matter under stress from a pole decomposition analysis of the
effective mass. We extract a complex-valued characteristic frequency
which displays scaling $|\omega^*(\sigma)|\sim\sigma^{\Omega'}$ with
vanishing stress $\sigma$ for a variety of granular systems.  The critical
exponent is smaller than that predicted by mean-field theory opening new
challenges to explain the exponent for frictional and dissipative granular matter.  Our
results shed light on the anomalous behavior of stress-dependent
acoustics and attenuation in granular materials near the jamming
transition.

\end{abstract}
\maketitle

\section{Introduction}

In granular media, a jammed system results if the particles pack
together at high enough density such that all the particles are
touching their neighbors \cite{jamming}.  As a result, the material
undergoes a jamming transition above which the system remains arrested
in a jammed state and is able to withstand a sufficiently small
applied stress. The jamming transition occurs at a particularly subtle
point where the particles have no redundant mechanical constraints,
i.e. they are isostatic \cite{alexander,moukarzel,makse,ohern1}.
The system is a fragile solid that is highly susceptible to external
perturbations. For instance, cutting a set of particle contacts
creates floppy modes which necessitate global rearrangement to
rebalance the system
\cite{alexander,moukarzel,Matthieu2005,Wyart2005,liu-review}.

This critical point has important consequences for the mechanical
properties of granular matter. In particular, there exists an excess
of low-frequency modes above that expected from the Debye scaling of
phonons in elastic solids \cite{ohern,silbert2005}. When the system is
jammed, a band of excess modes exists for $\omega > \omega_D^*$,
$\omega_D^*$ is a characteristic frequency that vanishes as the
jamming transition is approached \cite{silbert2005} [the subscript
  indicates that this frequency determines the limit of validity of
  Debye scaling in the density of states]. For $\omega < \omega_D^*$
one has the usual Debye density of states.  The system is argued to be
critical with a diverging correlation length $\ell^*$
\cite{Wyart2005} scaling as the inverse of the characteristic
frequency.
The excess modes distinguish an isostatic critical state at scales
below $\ell^*$ from an elastic solid at scales larger than $\ell^*$
\cite{Matthieu2005,Wyart2005}. The characteristic frequency follows
scaling law
\cite{ohern,silbert2005,liu-review,Matthieu2005,Wyart2005,friction1,xu}:
\begin{equation}
\omega_D^* \sim (\phi-\phi_c)^\Omega \sim \sigma^{\Omega'},
\label{critical}
\end{equation}
with external stress, $\sigma$, or volume fraction, $\phi$, measured
with respect to the jamming point $\Delta \phi = \phi - \phi_c$ (eg
random close packing or point J \cite{ohern,makse,ohern1}).

The static properties of jammed systems are also characterized by
scaling laws. In particular, the average coordination number scales as
\begin{equation}
Z-Z_c \sim (\phi - \phi_c)^{1/2},
\label{Z}
\end{equation} 
with scaling exponent approximately independent of friction
\cite{makse,ohern1,ohern,liu-review,friction1,zhang,Matthieu2005}. Here,
$\Delta Z = Z - Z_c$ measures the excess contacts from the minimal
isostatic coordination number for frictionless grains $Z_c=6$, or a
critical value $4\le Z_c \le 6$ for frictional grains in 3D
\cite{friction1,zhang,brujic-prl}.  The constitutive law between stress and
volume fraction is
\begin{equation}
\sigma
\sim (\phi-\phi_c)^\alpha,
\end{equation}
 where $\alpha=3/2$ for Hertzian spheres and $\alpha = 1$ for linear
 springs. Thus, $\Omega = \Omega' \alpha$.



To compare across systems with different constitutive laws, we
separate the trivial stress dependence of $\alpha$ from the
non-trivial structural dependence characterized by exponent $\delta$:
\begin{equation}
\omega_D^* \sim \Delta\phi^{(\alpha-1)/2} \Delta\phi^\delta,
\end{equation}
with $\delta = \Omega+(1-\alpha)/2$.  Theory
\cite{Wyart2005,Matthieu2005} has found using a variational argument
of boundary contact removal process for a system of particles with
linear constitutive law ($\alpha=1)$ 
\begin{equation}
\omega_D^*(\Delta Z) \sim \Delta Z, \,\,\,\,\,\, \mbox{($\alpha=1$,
  mean-field scaling)},
\label{the}
\end{equation}
 yielding a prediction $\delta_{\rm th}=1/2$, using Eq. (\ref{Z}),
 which agrees with simulations of frictionless particles
 \cite{silbert2005}.

So far, the scaling behavior of Eq. (\ref{critical}) has not been
tested by direct experimental measurements on zero-temperature
granular matter.  Indeed, experiments cannot calculate the density of
states by directly measuring the interparticle potential as done in
simulations \cite{ohern,silbert2005}. Here, we employ a novel {\it
  dynamical} measurement to measure the normal modes.

Previous experiments focused on thermally agitated
colloidal glasses and supercooled liquid systems
\cite{kurchan,Chen2010,Widmer2008}.
In contrast, granular materials are athermal (the energy necessary to
displace a grain is much larger than the thermal energy $\sim 10^{14}$
times) and friction dominated.  Moreover, Eq. (\ref{critical}) was
derived for non-frictional non-dissipative systems
\cite{Matthieu2005,Wyart2005}. In the case of real granular media,
dissipation plays a key role in governing the dynamics
\cite{JValenza2012}.  In this work, we develop an experimental study
of the stress-dependent normal modes frequencies for the case of
granular matter.  We demonstrate that the normal mode spectrum of a
finite sized granular system can be determined through a pole
decomposition of the frequency-dependent effective mass
\cite{Chaur2009,JValenza2009,JValenza2012}.  The results of this
analysis supports the scaling hypothesis Eq. (\ref{critical}) in a
realistic athermal and dissipative medium.  However, the exponents
characterizing this granular medium are smaller than those predicted
by theory Eq. (\ref{the}) \cite{Wyart2005,Matthieu2005}.

\section{Experiments}

We employ a dynamical effective mass measurement to obtain the normal
modes which complements other measures of the normal modes obtained
from the Fourier transform of the velocity autocorrelation function
and the eigenvalues of the displacement correlation matrix
\cite{ohern,silbert2005,xu,friction1,liu-review}.  Specifically, we
measure the frequency dependent effective mass, $\tilde{M}(\omega)$,
introduced previously in \cite{Chaur2009,JValenza2009,JValenza2012},
of a granular medium held in a cup under varying stress.

The full experimental program has been developed in
\cite{Chaur2009,JValenza2009,JValenza2012} and it is described in
detail in Appendix \ref{setup}.  Here we refer to the most
important aspects of the experiments and the corresponding set of
predictions.  The stress-dependent scaling behavior of the normal mode
spectrum of the granular system can be determined through a pole
decomposition of the frequency-dependent effective mass
\cite{Chaur2009,JValenza2009,JValenza2012}. We use a variety of
granular systems consisting of particles of different sizes (ranging
from $\sim$150 microns to millimeters), different shapes (irregular
and spherical), different materials (tungsten, steel and glass), and
different damping conditions (particles coated with silicone oil and
uncoated).

An important factor in the dynamics of this system is the existence of
dissipative modes.
Accordingly, our initial experiments follow the procedure described in
reference \cite{JValenza2012} to lightly coat the tungsten particles
(nominal size $\sim$150 $\mu$m) with silicone oil (PDMS), which yields
a system with known dissipative properties--- later, we will show
results for uncoated particles. The weakly wet powder is poured into a
cup, and the cup is tapped to encourage the powder to settle and yield
a flat free surface, Fig. \ref{cappedCup}. Finally, a standard
mechanical compaction protocol is followed
\cite{brujic,edwards,briscoe,Baule} to prepare the system in a reversible
state which yields a reproducible $\tilde{M}(\omega)$.

Once loaded, the cup is subjected to a vertical sinusoidal vibration
at angular frequency $\omega$. The force, $\tilde{F}(\omega)$, is
measured by a force gauge mounted between the cup and the shaker, and
the acceleration, $\tilde{a}(\omega)$, is taken as the average of that
measured at two points on the opposite end of a single diameter.  The
effective mass of the granular medium is the causal response function
\cite{Chaur2009,JValenza2009,JValenza2012}:
\begin{equation}
  \tilde{M}(\omega)
= \frac{\tilde{F}(\omega)}{\tilde{a}(\omega)} - M_c,
\label{meff}
\end{equation}
where $M_c$ is the mass of the empty cup. Here, $\tilde{M}(\omega) =
M_1(\omega) \:+\: iM_2(\omega)$ is complex-valued and reflects the
partially in-phase and out-of-phase motion of the individual grains
relative to the cup motion. $M_1$ is indicative of the elastic, and
$M_2$ of the dissipative characteristics of the powder. Figures
\ref{experiment}a and b show $M_1(\omega)$ and $M_2(\omega)$ for
tungsten systems exposed to an incrementally increasing uniaxial
confining stress, $\sigma$.  In all cases, the lowest characteristic
frequency mode manifests itself as the sharpest resonance peak in
$M_2(\omega)$. This peak and the other, lower amplitude, modes in the
system are characterized by a complex-valued normal mode frequency,
$\omega_n$. Roughly speaking, Re$[\omega_n]$ is given by the position
of the frequency mode, and Im$[\omega_n]$ is half of the full width at
half max of the frequency mode.

To accurately infer $\omega_n$ from the data in Fig. \ref{experiment}
we analyze $\tilde{M}(\omega)$ within the context of a theory of
damped and frictional contact forces \cite{JValenza2012}, which
generalizes the results of previous analyses
\cite{Matthieu2005,Wyart2005}. While previous theories have considered
central force systems \cite{Matthieu2005,Wyart2005}, interpreting the
dynamic response of our experiments necessitates a formalism that
accounts for translational and rotational degree of freedom and damped
modes. We develop such a formalism next.



\section{Theory}

\subsection{Interparticle force-law}
\label{theory_a}

In general the granular medium is modeled as a set of grains held in a
rigid cup. The theory is valid for any linearized set of contact
forces. Without loss of generality, below we define it for the
specific case of Hertz-Mindlin contact forces. The normal force
between any two contacting particles with radius $R$ is \cite{brilliantov,wolf}:
\begin{equation}
F_n=\frac{2}{3}k_{n}R^{1/2}x_{ij}^{3/2},
\label{n}
\end{equation}
where the normal deformation (1/2 the overlap between the spheres)
between the neighboring grains is $x_{ij}=\frac{1}{2}[2
  R-|\mathbf{x}_i-\mathbf{x}_j|]$, $\mathbf{x}_{i,j}$ are the position
vectors, and $k_n$ is the normal spring constant. The latter is
defined in terms of the corresponding material properties. The normal
elastic constant is
\begin{equation}
k_n= 4 G_g / (1-\nu_g),
\label{k-n}
\end{equation}
where $G_g$ is the shear modulus, and $\nu_g$ is the Poisson's ratio
of the material from which the grains are made.

The tangential force between neighboring grains in contact is
\cite{wolf}:
\begin{equation}
\Delta F_t = k_t (R x_{ij})^{1/2} \Delta s,
\label{t}
\end{equation}
where 
\begin{equation}
\label{k-t}
k_t = 8 G_g / (2-\nu_g),
\end{equation} is the tangential spring constant, and
the variable $s$ is defined such that the relative shear displacement
between the two grain centers is $2s$. Finally, Coulomb friction with
interparticle friction coefficient $\mu$ imposes $F_t \le \mu F_n$ at
every contact.

From the definition of particle interactions, the elastic spring
constant tensor is written in terms of the normal ($N$) and transverse
($T$) stiffnesses as:

\begin{equation}
\mathbf{K}_{ij}= k^N(x_{ij})\mathbf{\hat{d}}_{ij}\mathbf{\hat{d}}_{ij}
+ k^T(x_{ij}) [\mathbf{I} -
  \mathbf{\hat{d}}_{ij}\mathbf{\hat{d}}_{ij}],
\label{elastic}
\end{equation}
where $\mathbf{\hat{d}}_{ij}$ denotes the direction of the normal
displacement and we use the dyadic notation. Since we are interested
in infinitesimal displacements, without loss of generality we use a
linearized version of the Hertz-Mindlin force law Eqs. (\ref{n}) and
(\ref{t}) and define the elastic stiffnesses as:
\begin{equation}\label{Hertz}
k^N(x_{ij}) =k_n R^{1/2} x_{ij}^{1/2},
\end{equation}
and
\begin{equation}
k^T(x_{ij}) =k_t R^{1/2} x_{ij}^{1/2},
\label{Mindlin}
\end{equation}
where $k_n$ is the normal spring constant defined in terms of the
corresponding material properties above. That is, we use the
Hertz-Mindlin force law linearized about the static value of the
normal compression of contacts.  The resulting elastic stiffness is
simply the slope evaluated at $x_{ij}$.
The tangential force between neighboring grains in contact is defined
in terms of the tangential spring constant $k_t$ above.



The dissipative properties of the particles are defined as follows.
The definition of the damping tensor $\mathbf{B}_{ij}$ involves a
normal damping and a tangential damping in analogy to the definition
of the elastic matrix Eq. (\ref{elastic}). A representation of the
damping matrix in terms of the first order Taylor expansion gives the
following form which is an extension of the original Hertz approach
assuming the material to be viscoelastic instead of elastic. The form
of the damping tensor $\mathbf{B}_{ij}$ has been calculated by
Kuwabara and Kono \cite{kono} and Brilliantov {\it et al.}
\cite{brilliantov} (see also \cite{wolf} for a review):

\begin{equation}\label{Ndamp1}
\mathbf{B}_{ij}=\gamma^N(x_{ij})\mathbf{\hat{d}}_{ij}\mathbf{\hat{d}}_{ij}
+ \gamma^T(x_{ij}) [\mathbf{I} -
  \mathbf{\hat{d}}_{ij}\mathbf{\hat{d}}_{ij}].
\end{equation}
If the particles, $i$ and $j$, are not in contact
($|\mathbf{x}_j-\mathbf{x}_i|>2R$), we set both $\mathbf{K}_{ij}=0$
$\mathbf{B}_{ij}=0$.

Both damping constants, normal $\gamma^N(x_{ij})$ and tangential
$\gamma^T(x_{ij})$, are proportional to the respective elastic
constants in the elastic counterparts $k^N(x_{ij})$ and
$k^T(x_{ij})$. This is seen by following the calculations of
Brilliantov \cite{brilliantov} for the dissipative force between two
contacting particles. The total force acting between viscoelastic
particles can be derived from the total stress tensor taking into
account the elastic and dissipative parts (see Landau for details
\cite{landau}):
\begin{equation}
\hat{\sigma} = \hat{\sigma}_{(el)} + \hat{\sigma}_{(dis)}.
\end{equation}
The calculations are simplified since the elastic and dissipative
parts of the stress tensor are related in the quasi-static limit
\cite{brilliantov,landau}:
\begin{equation}
\hat{\sigma}_{(dis)} = \dot{x}_{ij} \frac{\partial}{\partial x_{ij}}
\hat{\sigma}_{(el)}.
\end{equation}
This leads to a dissipative force of the form:
\begin{equation}
F^N_{dis} = \xi k_n R^{1/2} x_{ij}^{1/2} \dot{x}_{ij},
\label{fn_diss}
\end{equation}
where $\xi$ is the damping parameter with unit of time and related to
the elastic and viscoelastic constant of the material from which the
particles are made. From \cite{landau} we have:
\begin{equation}
\xi = \frac{1}{2} \frac{(3\eta_2-\eta_1)^2}{(3\eta_2+2\eta_1)}
\Big[\frac{(1+\nu_g) (1-2\nu_g)}{2G_g}\Big],
\end{equation}
where $\eta_1$ and $\eta_2$ are the viscous constants of the material
of the particles (see Eq. (23) in \cite{brilliantov}).  Comparing with
the Hertzian counterpart Eq. (\ref{n}) we have the formal relation for
the total force in the normal direction:
\begin{equation}
F^N_{tot} = const \Big(x_{ij}^{3/2} + \xi x_{ij}^{1/2} \dot{x}_{ij}\Big).
\end{equation}
A similar derivation holds for the tangential components, for which we
have \cite{brilliantov}:
\begin{equation}
F^T_{dis} = \xi k_t R^{1/2} x_{ij}^{1/2} \dot{s}.
\label{ft_diss}
\end{equation}
These considerations leads to Eq. (\ref{Ndamp1}) with
\begin{equation}
\gamma^N(x_{ij}) = \xi k^N(x_{ij}),
\end{equation}
 and
\begin{equation}
\gamma^T(x_{ij}) = \xi k^T(x_{ij}).
\end{equation} 
Combining these equations we arrive at:

\begin{equation}
\mathbf{B}_{ij} = \xi \mathbf{K}_{ij},
\label{BKanalysis}
\end{equation}
which is used later to understand the trajectories of the normal mode
frequencies in the complex plane, considered as functions of $\xi$:
$\omega_n = \omega_n(\xi)$.



\begin{figure}
(a) \centering{ \includegraphics[width=0.5\columnwidth]{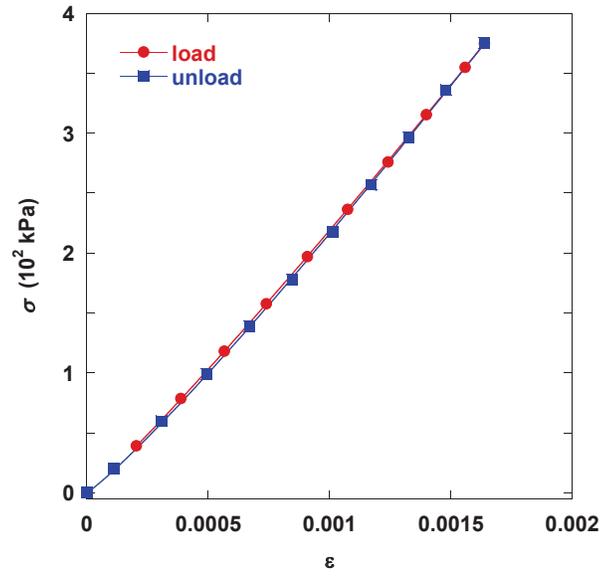}}
  \centering{
    \includegraphics[width=0.5\columnwidth]{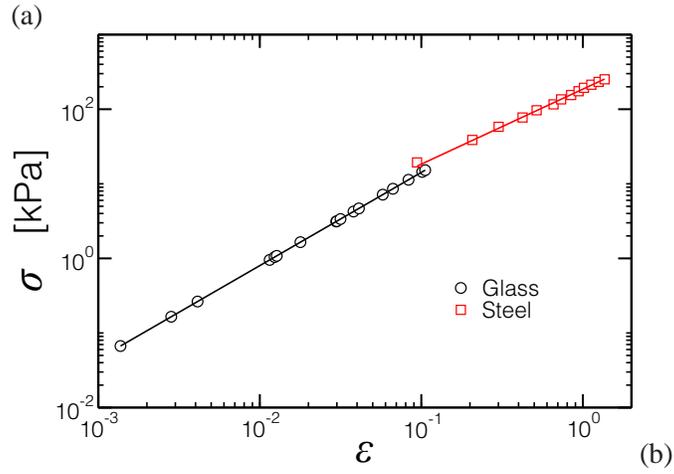}}
  (b)\caption{(Color online) Force constitutive law.  We measure the axial stress
    vs. strain response of (a) the tungsten powder confined in the cup
    and find an approximate power-law with exponent $1.15\pm0.01$,
    averaged over load and unload, (b) the packing of glass beads and
    steel balls with exponents $\alpha$ given in Table \ref{table2}.}
\label{Fvd}
\end{figure}

The constitutive contact laws expressed by the Hertz-Mindlin theory
imply the trivial scaling between the stress and strain (or volume
fraction) under compression \cite{HMakse}:
\begin{equation}
\sigma \sim \epsilon^{\alpha},
\label{hertzNL}
\end{equation}
with $\alpha=3/2$ for Hertz contact force law, Eq. (\ref{n}), $\sigma$
is the uniaxial stress and $\epsilon$ is the strain.  When the
constitutive particles follow other force law, the exponent $\alpha$
is modified accordingly.  In fact, $\alpha=3/2$ is expected for a
collection of monodisperse spherical Hertzian grains, which is not an
accurate description for the system we study here. Therefore, we use a
mechanical testing machine to measure the stress-strain relationship
exhibited by the tungsten powder and the system of glass and steel
beads confined in the cup. To perform this characterization, we sit a
solid stainless steel plunger on top of the powder. As demonstrated in
Fig. \ref{Fvd}a, we find a weakly non-linear stress-strain response
of the tungsten powder. For the powder we use in this study, the power
law exponent always falls significantly below that predicted by
Eq. (\ref{hertzNL}), $\alpha=1.15\pm0.01$. We also test the
constitutive laws for the systems of glass beads and steel beads
(Fig. \ref{Fvd}b) obtaining the $\alpha$ exponent summarized in
Table \ref{table2}.

\subsection{Theory of normal modes in a dissipative medium}
\label{theory_b}

In what follows, $\mathbf{x}_i$, and $\mathbf{u}_i$ denote the
equilibrium position, and displacement from equilibrium, respectively,
of the $i$-th particle. The variables $\theta_i$ represent the
librational angles of the i-th particle. In addition, $\mathbf{W}=W
\mathbf{\hat {z}}$ represents the displacement of the cup wall in
the $z$-direction. In the associated experiment, $\mathbf{W}=W
\mathbf{\hat {z}}$ is 1 micron, at least three orders of magnitude
smaller than $R$, therefore, all $\mathbf{u}_i$ are infinitesimal and
the linear equation of motion for the $i$-th particle with mass $m$
is:
\begin{multline}\label{NewtonEq}
-m\omega^2\mathbf{u}_i=\mathbf{K'}_{i\omega}[\mathbf{u}_i+\theta_i
  \times\mathbf{\hat{d}}_{iw}
  -\mathbf{W}] + \\
\sum_j\mathbf{K'}_{ij}[\mathbf{u}_j-\mathbf{u}_i
  +\theta_j \times\mathbf{\hat{d}}_{ji} - \theta_i
  \times\mathbf{\hat{d}}_{ij} ],
\end{multline}
where, $\mathbf{d}_{ij}=(1/2)(\mathbf{x}_i-\mathbf{x}_j)$ for the case
when neighboring particles are identical spheres, and the subscript,
$w$ indicates a particle interacting with the cup wall.
The equations of motion for the angular variables are:
\begin{multline}\label{Newtonang}
- \omega^2\mathbf{I}_i \dot \theta_i = - \mathbf{d}_{iw} \times
\mathbf{K'}_{i\omega}[\mathbf{u}_i+\theta_i
  \times\mathbf{\hat{d}}_{iw} -\mathbf{W}]+\\
\sum_j \mathbf{d}_{ij}
\times \mathbf{K'}_{ij}[\mathbf{u}_j-\mathbf{u}_i +\theta_j
  \times\mathbf{\hat{d}}_{ji} - \theta_i \times\mathbf{\hat{d}}_{ij}
].
\end{multline}

The stiffness matrix $\mathbf{K'}_{ij}$ modulates the elastic and
viscous interactions between neighboring particles and is defined in
terms of the elastic matrix $\mathbf{K}_{ij}$ and the damping matrix
$\mathbf{B}_{ij}$ as defined above:
\begin{equation}
\mathbf{K'}_{ij}= \mathbf{K}_{ij} -i \omega \mathbf{B}_{ij}.
\end{equation}
Similarly, $\mathbf{K'}_{iw}$ describes the elastic and viscous
interaction between a particle and the wall of the cup; it is zero
for particles not in contact with the wall.  Next we write a compact
form of these equations of motion, Eqs. (\ref{NewtonEq}) and
(\ref{Newtonang}).


The linearized equation of motion for the $i-$th particle in the system
of $N$ particles with mass $m$ can be succinctly written as:
\begin{equation}\label{h}
H_{ij}(\omega) u_j = K_{i\omega} W, \,\,\,\,\,\, \mbox{$i,j = 1 :
  6N$},
\end{equation}
where the dynamical matrix is:
\begin{equation}
H_{ij}(\omega)=- m \delta_{ij} \omega^2 - i \omega B_{ij} + K_{ij}.
\label{matrix}
\end{equation}
Each term in $H_{ij}$ accounts for the inertial, dissipative and
elastic interactions at the contact between particles $i$ and $j$,
respectively. 
The vector $\{u_j\}$ represents the set of $3N$ particle displacements
and $3N$ particle rotations, and $K_{i\omega}$ is the generalized
spring constant connecting a particle to the walls of the cup which
moves oscillatory in the $z$ direction with amplitude $W$.  The
effective mass is obtained by inverting the matrix $\mathbf{H}$
\cite{JValenza2012}:
\begin{equation}\label{effmass}
\tilde{M}(\omega)=m
    [H^{-1}(\omega)]_{ij} K_{j\omega}.
\end{equation}
This result formalizes the relation between the resonance peaks
in the effective mass and the normal mode frequency spectrum. The
peaks observed in $\tilde{M}(\omega)$ (Fig. \ref{experiment}) are due
to the set of normal modes, $e_j^n$, that are a solution to
Eq. (\ref{h}) when there is no forcing by the cup, $W=0$, i.e.,
\begin{equation}
H_{ij}(\omega_n) e_j^n = 0.
\label{eigen}
\end{equation}
The normal modes are those eigenvectors of $\mathbf{H}$ for which the
corresponding eigenvalue is zero, and they occur at the complex-valued
frequencies, $\omega_n$.

The normal mode frequencies, $\omega_n$, are the non-trivial solutions
of Eq. (\ref{eigen}), in which the eigenvalue is 0. Regardless the
properties of the matrices $\mathbf{K,B}$ it is a rigorous result that
for values of $\xi$ below a critical and finite value, $\xi_m$, all
the modes are underdamped \cite{InmanAndry}, meaning
$\Re(\omega_n)\neq 0\:\:\forall \;n\;$.  Similarly, there is another
finite, critical value, $\xi_M$, such that when $\xi\ge \xi_M$ all the
modes are overdamped \cite{Bhaskar}: $\Re(\omega_n)\equiv 0
\;\;\forall \;n$.  As $\xi$ is increased from $\xi_m$ to $\xi_M$ each
branch becomes critically damped at the values $\xi_c$ viz:
$\omega_n(\xi_c) = i\lambda_n\:\:\:n = 1, 2, ...$.  All normal modes
have the same functional dependence on $\xi$ in the vicinity of such a
critical point.  Let $D(\lambda, \xi) = det\{H_{ij}(\omega= i\lambda,
\xi)\}$.  In the vicinity of the critical damping point $\lambda =
\lambda_{nc},\;\xi = \xi_{nc}$ we may expand $D$ in a Taylors series:
\begin{equation}\label{Taylor}
\begin{array}{r} 
D(\lambda,\xi) = a_n(\lambda - \lambda_{nc})
+ b_n(\xi -
\xi_{nc}) + d_n(\lambda - \lambda_{nc})^2 +\\
\\
 e_n(\lambda - \lambda_{nc})(\xi - \xi_{nc}) + f_n
(\xi - \xi_{nc})^2\;+\;\ldots\end{array}\:\:\:,
\end{equation} 
where 
\begin{equation}\label{Taylor2}
\begin{array}{c} 
a_n =
\frac{\partial D}{\partial \lambda}|_{(\lambda_{nc},\xi_{nc})}\:\:\:b_n = \frac{\partial
D}{\partial
\xi}\mid_{(\lambda_{nc},\xi_{nc})}\\
\\
d_n = (1/2)\frac{\partial^2 D}{\partial \lambda^2}|_{(\lambda_{nc},\xi_{nc})} \:\:\:e_n =
\frac{\partial^2 D}{\partial \lambda \partial \xi}|_{(\lambda_{nc},\xi_{nc})}\\
\\ f_n = (1/2)\frac{\partial^2 D}{\partial \xi^2}|_{(\lambda_{nc},\xi_{nc})}
\end{array}\end{equation} 
The coefficients $a_n, b_n, \ldots , f_n$ are all real-valued because
$\{a_i^j\}, \lambda_{nc}$ and $\xi_{nc}$ are all real-valued.
$\lambda_{nc}$ is a double root of $D$ because it represents the
coalescence of two distinct complex-conjugate roots in the limit $\xi
\rightarrow \xi_{nc}^{-}$. Accordingly, $a_n \equiv 0$.  One may solve
for the roots of Eq. (\ref{Taylor}):\begin{equation}\label{crit}
  \omega_n = i\lambda_n = i\lambda_{nc} \pm i g_n(\xi -
  \xi_{nc})^{1/2} + \mathcal{O}(\xi -
  \xi_{nc})^{+1}\:\:\:,\end{equation} where $g_n = \sqrt{-b_n/d_n}$ is
real-valued because $\lambda_n$ is real-valued when $\xi>\xi_{nc}$.

Since $\mathbf{H}$ is a $6N \times 6N$ matrix, there are 12$N$ normal
modes in this model. In practice, the effective mass can be understood
in terms of a subset of these normal modes, which correspond to the
modes located in the complex region of interest $[-\omega_M \le {\rm
    Re} [\omega_n] \le +\omega_M\;:\:\:-\omega_M \le {\rm Im}
  [\omega_n], \le 0]$, where $\omega_M$ is the maximal frequency
measured in the experiment.  The frequency dependence of
$\tilde{M}(\omega)$ can be expressed in terms of this subset of normal
modes through a pole decomposition \cite{JValenza2012} as described in
the next section.  

Thus, this set of equations
allow for an experimental measurement of the normal mode frequencies
directly from the effective mass.

\section{Pole decomposition}
\label{pole}

The matrix $\mathbf {H}$ is complex-valued, frequency-dependent and
symmetric.  A non-Hermitian matrix like $\mathbf {H}$ may lack, in
general, the property that its eigenvectors form a complete
orthonormal basis \cite{19}. However, according to a recent theorem by
Tzeng and Wu \cite{20}, there still exists $6N$ orthonormal vectors
which satisfy the following modified eigenvector problem:
\begin{equation}
H_{ij} e_j^n = \lambda^n e_i^{n*},
\end{equation}
where the eigenvalues $\lambda^n$ are complex valued and the asterisk
denotes conjugation. A normal mode of the system is a solution to the
set of Eqs. (\ref{NewtonEq}) and (\ref{Newtonang}) in which there is
no forcing by the cup: $W=0$. The corresponding displacements or
rotations are the eigenvectors of $H$ whose eigenvalues are zero, i.e.,
$\lambda^n(\omega_n) = 0$. If we expand to first order:
$\lambda^n(\omega) = \alpha^n(\omega - \omega_n) + O(\omega -
\omega_n)^2$, and introduce this expansion into Eq. (\ref{effmass}),
we obtain the pole expansion:
\begin{equation}
  \tilde{M}(\omega) = \sum_n \frac{A_n}{\omega-\omega_n},
\label{pd}
\end{equation}
with the residues given by
\begin{equation}
A_n = \frac{m e^n_i(\omega_n) e_j^n(\omega_n) K_{jw}(\omega_n)}{\alpha^n}.
\end{equation}

In reference \cite{JValenza2012} it was demonstrated that the sum
Eq. (\ref{pd}), is an exact expression for the effective mass. This
allows us to use the rational function approximation, $\tilde{M}_{RF}$
discussed in the following section to extract the normal mode
frequencies from the effective mass.  The matrix $\mathbf{A_n}$ are
the residues of the poles representing the strength of each resonance
$\omega_n$. Using this expansion, which can be proved to be exact
\cite{JValenza2012}, the set of normal modes and residues $\{
\omega_m,A_m\}$ can be determined from the experimental data of
Fig. \ref{experiment} via a search of all zeros of the rational
function $M^{-1}$ analytically continued to the complex plane. We
describe this procedure in the next Section \ref{rf}.

\section{Rational function decomposition to extract normal modes}
\label{rf}

We use the measured effective mass to extract the normal modes via the
pole decomposition described by Eq. (\ref{pd}). Here, we describe the
numerical procedure for generating the rational function interpolation
of our data $\tilde{M}_{RF}$, which we use to determine the poles and
the residues in our system.

In the experiments, we measure the effective mass
$\tilde{M}(\omega_i^e)$ at a series of discrete frequencies $e=1:
1490$ from 100 Hz to a maximum frequency of 15 kHz. Using the
reflection property,
$\tilde{M}(-\omega_i^{e*})=\tilde{M}^*(\omega_i^e)$, we extend the
experimental data to negative real-valued frequencies, and we set
$\tilde{M}(0)=M_0$, the static mass of the powder. So we are working
with 2981 data points on the real frequency axis.

To search for the normal mode frequencies we require a complex valued
function that passes through all the data points on the real axis. To
this end we employ the Bulirsch-Stoer algorithm
\cite{Bulirsch-Stoer-algorithm} to determine a rational interpolation
$R_{i(i+1)\ldots(i+m)}(x)$ of our data
$(x_i,y_i),\ldots,(x_{i+m},y_{i+m}).$. Typically
\cite{Numerical_Recipes}:
\begin{equation}\label{RF}
R_{i(i+1)\ldots(i+m)}(x)=\frac{p_0+p_1(x)+\ldots +p_{\mu}x^{\mu}}{q_0+q_1(x)+\ldots+ q_{\nu}x^{\nu}},
\end{equation}
where the coefficients of the numerator and denominator can be complex
numbers.
We choose to use the rational function in the recurring form:
\begin{equation}
R_{i(i+1)\ldots(i+m)}(x)=\frac{R_{(i+1)\ldots(i+m)}(x)-R_{i\ldots(i+m-1)}(x)}{(\frac{x-x_i}{x-x_{i+m}})(1-\frac{R_{(i+1)\ldots(i+m)}(x)-R_{i\ldots(i+m-1)}(x)}{R_{(i+1)\ldots(i+m)}(x)-R_{(i+1)\ldots(i+m-1)}(x)})},
\end{equation}
where, the starting points are $R_i(x)=y_i,$ $R_{i-1}(x)=0.$

The poles $\omega_n$ correspond to
$M_{RF}(\omega_n)=\infty$. Therefore, we identify the poles using the
condition outlined by Eq. (\ref{complexPoles}),
$\frac{1}{M_{RF}(\omega_1)}=0$. We employ Muller's method
\cite{Numerical_Recipes} to identify the poles in the complex
plane. Once a normal mode frequency is identified, $\omega_1$, we use
Eq. (\ref{pd}),
to determine the corresponding residue $A_1$,
$A_1=\lim_{\omega\rightarrow\omega_1}
(\omega-\omega_1)M_{RF}(\omega).$ Using the reflection property, we
obtain the corresponding symmetric pole and residue which are
$-\omega_1^*$ and $-A_1^*.$

The main resonance peak in $\tilde{M}(\omega)$ is often very large
compared to the rest of the peaks, which makes it simple to find the
first pole, but increasingly difficult to find the remaining poles by
Muller's method. Thus, to find the next pole we use the difference
between the original effective mass and the effective mass given by
Eq. (\ref{pd}). That is, we define a new set of data points
$\{\omega_i^e, M_i^2\}$ by
\begin{equation}
M_i^2(\omega_i^e)=M_i(\omega_i^e)-\frac{A_1}{\omega_i^e-\omega_1}-\frac{-A_1^*}{\omega_i^e-(-\omega_1^*)}.
\end{equation}
This iterative process is repeated to identify all of the poles that
make a notable contribution to the measured effective mass. As a test
of the accuracy of this approach, we compare the real data to that
from Eq. (\ref{pd}) utilizing all of poles and residues as shown in
Fig. \ref{fitting}. The difference between the data and the fitting
  is negligible indicating that the relevant poles in the frequency
  range of measurement have been identified.
The other systems at different stress behave similarly.

 As explained above, there are $12N$ normal modes. The modes that we
 observe in the effective mass are those with the largest residues in
 the pole decomposition as expressed by Eq. (\ref{pd}). While it is
 visually apparent that only a few modes appear in the experimental
 effective mass, like in Fig. \ref{experiment}, when we subtract the
 pole contribution of the principal mode, then a finer structure
 appears, as seen in Fig. \ref{fitting}. The curve called Remainder in
 Fig. \ref{fitting} appears with many (small) peaks signaling the
 existence of many more modes in the system. The fact that these modes
 are visible only after subtracting the first modes, is because their
 residues are very small and therefore do not contribute much to the
 effective mass. Thus, the effective mass is sensitive to the most
 important extended modes in the system as given by the visible
 resonance peaks. These modes are extended and define the correlation
 length of the system. The remaining modes are still part of the
 effective mass but their contribution is small.

To locate the remaining modes, we repeat the process of subtracting
the pole contribution from Eq. (\ref{pd}) after we locate the largest
pole in the remainder signal. In this way, we keep locating all the
modes in the system. We estimate that we have approximately 1 million
particles in the tungsten system. While it is not realistic to expect
to find all the 6 million modes by this method, the pole decomposition
provides a large number of the most important modes that define the
effective mass in the region of observation. Indeed, there could be
other modes that are outside the frequency domain of measurement and
cannot be measured.  Beyond this, the only limitation to locate the
modes is the resolution of the signal obtained as a remainder after
subtracting each pole from Eq. (\ref{pd}).


\section{Interpretation of experimental data}
\label{interpretation}

In practice, we fit a complex valued rational function to our
experimental data, $\tilde{M}_{RF}(\omega)$ to determine the set of
perceivable normal modes and residues $\{\omega_n,A_n\}$
(Fig. \ref{experiment}). Utilizing $\tilde{M}_{RF}(\omega)$ the normal
modes are identified by the criteria
of Section \ref{rf}:
\begin{equation}
 \frac{1}{\tilde{M}_{RF}(\omega_n)} = 0.
\label{complexPoles}
\end{equation}

The locations of the complex-valued normal modes in the $\omega$-plane
(Re$[\omega_n]$, Im$[\omega_n])$ are plotted in Fig. \ref{residues}
for the tungsten system at $\sigma=44.8$ kPa.
The poles are located relatively close to the real axis, indicating
that the system is underdamped, Im$[\omega_n]\ll$ Re$[\omega_n]$,
which is consistent with our previous experience \cite{JValenza2012}.
Furthermore, the data of Fig. \ref{residues} of the identified normal
modes frequencies follow, in average, a parabolic curve in the complex plane.
We have confirmed that this parabolic shape also holds for the other
stress levels, however the data is more noisy. This parabolic shape is
also confirmed in numerical simulations done in \cite{hu}.

This result provides the basis for interpreting the experimental data.
Indeed, the parabolic curve can be shown to be the result of a weakly
damped system with a commutative property of the dynamical matrices:
if the damping and stiffness matrices commute, $\mathbf{KB} =
\mathbf{BK}$, the trajectories can be approximated by parabolas 
for small damping.
As an illustrative example of this, let us investigate a consequence
of the approximation Eq. (\ref{BKanalysis}) between the elastic and
the damping matrices.


An important implication of Eq. (\ref{BKanalysis}) is that the
matrices $\mathbf{K}$ and $\mathbf{ B}$ commute: $\mathbf{KB} =
\mathbf{BK}$, and the set $\{e_j^n\}$ are the complete eigenvectors
for both matrices.  The normal modes in the damped system are exactly
the same as in the undamped case except that they now have
complex-valued frequencies, due to the attenuation.  In the presence
of damping, each mode exactly decouples.

 Expanding Eq.  (\ref{eigen}) we have,
\begin{equation}\label{DaveEq}
\Big [-m \delta_{ij} \omega_n^2- i \omega_n B_{ij} + K_{ij} \Big] e_j^n=0.
\end{equation}
When $\xi>$ 0 the normal mode frequencies are complex. If we
substitute the approximation Eq. (\ref{BKanalysis}) into
Eq. (\ref{DaveEq}), we have:

\begin{equation}\label{DaveEqNew}
\Big [-m \delta_{ij} \omega_n^2- i \omega_n \xi
    K_{ij} + K_{ij} \Big] e_j^n=0.
\end{equation}
The approximation Eq. (\ref{BKanalysis}) implies that $e_j^n$ is also
an eigenvector of $\mathbf{B}$, which permits us to simplify
Eq. (\ref{DaveEqNew}):
\begin{equation}\label{quadNorm}
- \omega_n^2-i \xi \omega_{n0}^2
  \omega_n  + \omega_{n0}^2 =0,
\end{equation}
where $\omega_{n0}$ are the normal modes of the system without
damping, $\xi=0$. Equation (\ref{quadNorm}) is a quadratic equation
with roots:
\begin{equation}\label{omega}
\omega_n=- i \frac{\xi}{2} \omega_{n0}^2 \pm
\omega_{n0} \sqrt{ 1 - \Big( \frac { \xi \omega_{n0} } { 2} \Big)^2}.
\end{equation}


For large enough damping, $\xi > \xi_c = 2 / \omega_{n0}$, these
normal modes are overdamped, and the corresponding frequencies are
purely imaginary. For weak damping, $\xi < \xi_c$, the modal
frequencies are damped oscillators. In this case:
\begin{equation}
\xi < \xi_c:\; |\omega_n(\xi)| =  \omega_{n0}.
\label{omega0}
\end{equation}

Thus, when Eq. (\ref{BKanalysis}), the underdamped modes have the
property that,
for a fixed $\xi \ll \xi_c$, the modes lie approximately on a parabola
given by:
\begin{equation}
\mbox{Im}[\omega_n] \approx - \frac {\xi}{2} \; \mbox{Re}[\omega_n]^2.
\label{quadratic}
\end{equation}

As seen in Fig. \ref{residues}, the modes exhibit a parabolic shape in
the $\omega_n$-plane, which suggests the validity of
Eq. (\ref{quadratic}) in average.  We fit Eq. (\ref{quadratic}) to the
modal frequencies obtained experimentally in Fig. \ref{residues}, and
find $\xi= (2.2\pm0.2)\times 10^{-5}$s$\ll\xi_c=2/\omega_{n0}$,
confirming that the system is weakly damped. We conclude that the
scaling of the undamped normal modes can be extracted via
Eq. (\ref{omega0}), i.e. by plotting the absolute value, $|\omega_n|$
against applied stress.


\section{Location of the missing modes}

\begin{figure}
\resizebox{6.5cm}{!}
          {\includegraphics{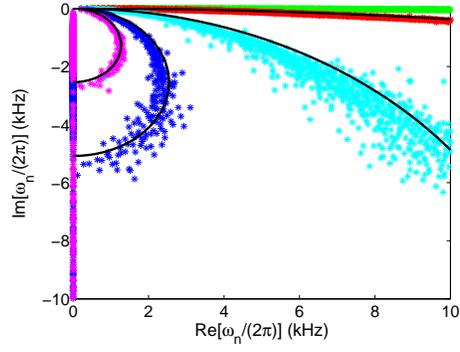}} \caption{(Color online) Computed
            complex-valued normal mode frequencies of a simulated
            system of 400 particles interacting via contact springs
            and dampers for 5 different values of the damping
            parameter.  From Ref. \cite{hu}. Although the matrices
            {\bf K} and {\bf B} do not commute in this case, the
            normal mode frequencies approximately follow the
            predictions of Eq. (\ref{circle_plot}), which is shown as
            a solid curve for different values of $\xi$.}
\label{cross-plot}
\end{figure}

At this point we would like to show how the results of our analysis
can give an indication of where, in the complex plane, the remaining
normal mode frequencies are located.  These missing normal mode
frequencies correspond to residues which are negligibly small in the
measured effective mass and are not visible in the effective mass
decomposition. Consider Eq. (\ref{omega}) which may be re-written as
\begin{equation}\label{circle_plot}
[\Re(\omega_n)]^2 + [\Im(\omega_n)+\xi^{-1}]^2 = \xi^{-2}
\end{equation}
Thus, if the matrices {\bf B} and {\bf K} commute all the normal mode
frequencies, for a fixed value of the damping parameter, $\xi$, lie on
a circle of radius $1/\xi$ centered on the point $-i/\xi$, as
discussed above.  In a separate effort of ours \cite{hu} we have
investigated the extent to which these results may hold true with
computer simulations.  We have performed molecular dynamic simulations
of ensembles of spherical grains in which we take the spring
constants, {\bf K}, to be given by the Hertz-Mindlin theory and we
have taken the damping constants, {\bf B}, to be the same for every
non-zero contact.  Although {\bf B} and {\bf K} do not commute in this
case, the computed normal mode frequencies are reasonably well
described by the predictions of Eq. (\ref{omega}) as we show in
Fig. \ref{cross-plot} which is reproduced from \cite{hu}.
Specifically, for each value of the damping parameter, $\xi$, the set
$\{\omega_n\}$ approximately lies on a circle whose radius decreases
as $\xi$ is increased, Eq. (\ref{circle_plot}).  The matrices {\bf B}
and {\bf K} are mostly zero-valued, except for grains which are
actually in contact with each other.  In this sense we may say that
they approximately commute; hence Eqs. (\ref{omega0}) and
(\ref{circle_plot}) are approximately true.

We suppose these features also apply to real data on real granular
media as we have already seen in Fig. \ref{residues}. In order to draw
some quantitative conclusions from our data we replot, in
Fig. \ref{New_fig}, some of our previously published results
\cite{JValenza2012} which were deduced from the measured effective
mass of tungsten granules lightly coated with viscous PDMS fluid. The
difference between the system of coated tungsten particles used in
Fig. \ref{residues} and Fig. \ref{New_fig} is that in the later we use
larger amount of silicone oil to coat the particles (80 mg as
indicated). Thus, the damping in the system is increased as shown in
\cite{JValenza2012}.  We have fit the data to Eq. (\ref{circle_plot})
using the cost function
\begin{equation}\label{cost_function}
\chi(R) = \sum_n\mid A_n\mid r_n^2\;[\theta_n - \theta(r_n;R)]^2\:\:\:.
\end{equation}
Here, $R=1/\xi$ is the radius of the circle and the resonant
frequencies are written in polar coordinates: $\omega_n = r_n
\exp{(i\theta_n)}$. $A_n$ are the residues of the poles in the
decomposition of the effective mass data, Eq. (\ref{pd}); their
magnitudes are indicated by the sizes of the symbols in
Fig. \ref{New_fig}. $\theta(r;R) = -\arcsin(r/2R)$ is the
representation of Eq. (\ref{circle_plot}) in polar coordinates.  The
results are shown in Fig. \ref{New_fig}(a).

Although the existing data cover only a small arc of the circle, we
feel that based on the results of our numerical simulations, and the
fact that the existing data do trend along the curve,
Eq. (\ref{circle_plot}), we may conclude that the underdamped
resonance frequencies which we are not able to directly locate with
our effective mass measurements lie roughly between the bounds of the
two dashed curves in Fig. \ref{New_fig}.  The overdamped resonance
frequencies will lie on the negative imaginary axis, of course.

\section{Stress-dependence of the characteristic frequency}

\begin{figure}
\centering{ \includegraphics[width=.9\columnwidth]{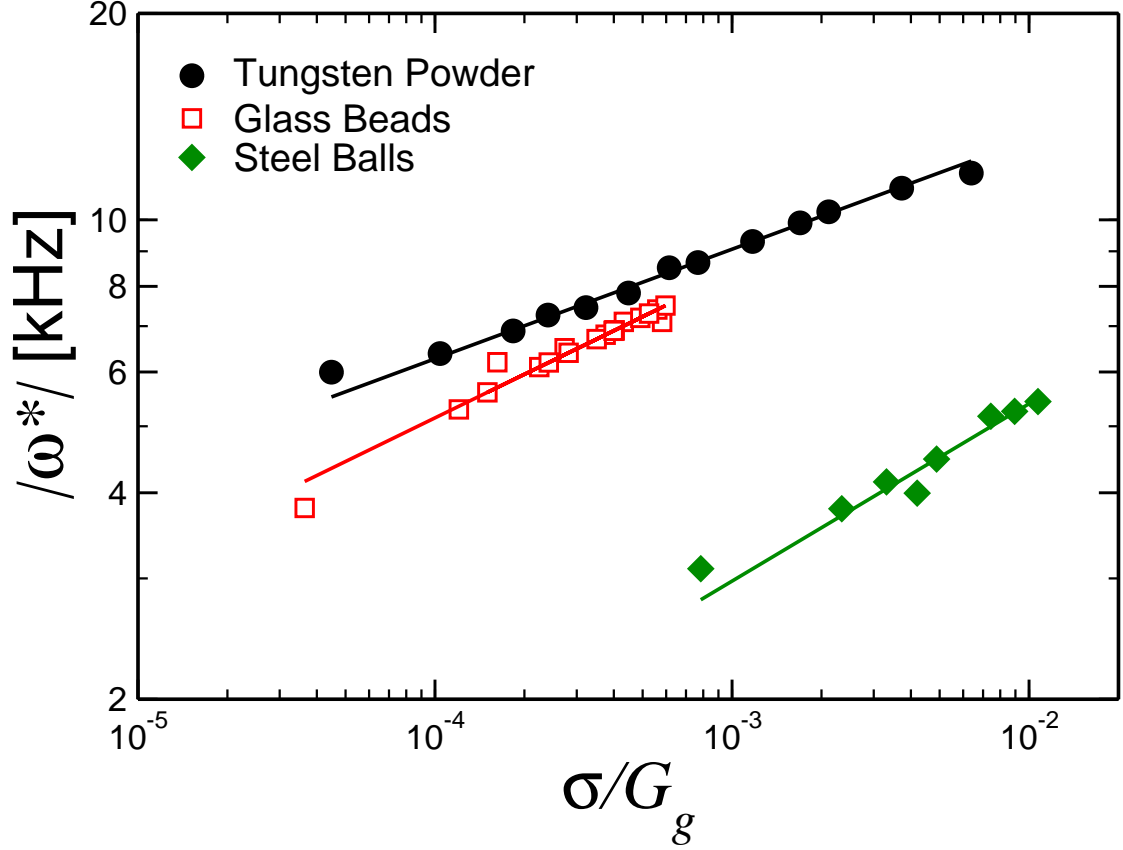}}
\caption{(Color online) Absolute value of the characteristic frequency,
  $|\omega^*(\sigma)|$, corresponding to the main resonance peak of
  the effective mass as a function of stress $\sigma$ rescaled by the
  shear modulus ($G_g$) of the material from which the particles are
  made. We plot
results for experimental systems composed of tungsten irregular
particles of $\sim 150 \mu$m coated with PDMS, uncoated spherical
glass beads of 1mm, uncoated steel balls of 2mm.
Solid lines are power-law fits with OLS estimator. Exponents are
reported in Table \ref{table2}.}
\label{scaling}
\end{figure}

The pole with the largest residue contributing to the resonance
frequency at the main peak of the effective mass defines the
characteristic frequency $|\omega^*(\sigma)|$. By the absolute value
we indicate that this is the characteristic frequency of the damped
modes obtained via the effective mass to distinguish it from the
undamped $\omega_D^*$ obtained from the Debye departure in the density
of states in Eq. (\ref{critical}). Figure \ref{scaling} shows
$|\omega^*(\sigma)|$ indicating a power-law dependence on $\sigma$
consistent with Eq. (\ref{critical}) for the tungsten powder. Using an
OLS estimator, we find $\Omega'=0.15\pm 0.02$ ($R^2 = 0.9949$).  More
specifically, the fitting yields:
\begin{equation}
|\omega^*(\sigma)| = (3.27\pm 0.03)\,
\sigma^{\Omega'},
\label{tungsten}
\end{equation}
with $\sigma$ measured in kPa and the frequency in kHz.

Using the stress vs strain curve in Fig. \ref{Fvd}a with $\alpha =
1.15 \pm 0.01$,
we find $\Omega = 0.17\pm 0.02$ and $\delta =
0.10\pm 0.02$ for the tungsten powder.


We test further the results with packings constituted by different
particle types. We use spherical particles made of glass beads (1mm
diameter) and steel balls (2mm diameter) without coating.  Figure
\ref{scaling} shows the scaling of $|\omega^*|$ with stress for
these systems and the exponents ($\Omega', \Omega, \delta, \alpha$) are
indicated in Table \ref{table2} (see Fig. \ref{Fvd}b for constitutive
law). We find for glass beads $\delta=0.14\pm0.02$ and for steel balls
$\delta = 0.13\pm 0.02$.

The exponent of the tungsten particles is smaller than the rest. We
hypothesize that irregularities in particle shape may add some
complexity in the modes not seen in spherical particles.  These
particles were obtained by fusing two or three irregular
(approximately cubic) particles with facets and angularities that may
introduce modes not seen in spherical particles coupling the
rotational and translational modes in an uncontrolled way.

\begin{figure}
\centering{  \includegraphics[width=0.9\columnwidth]{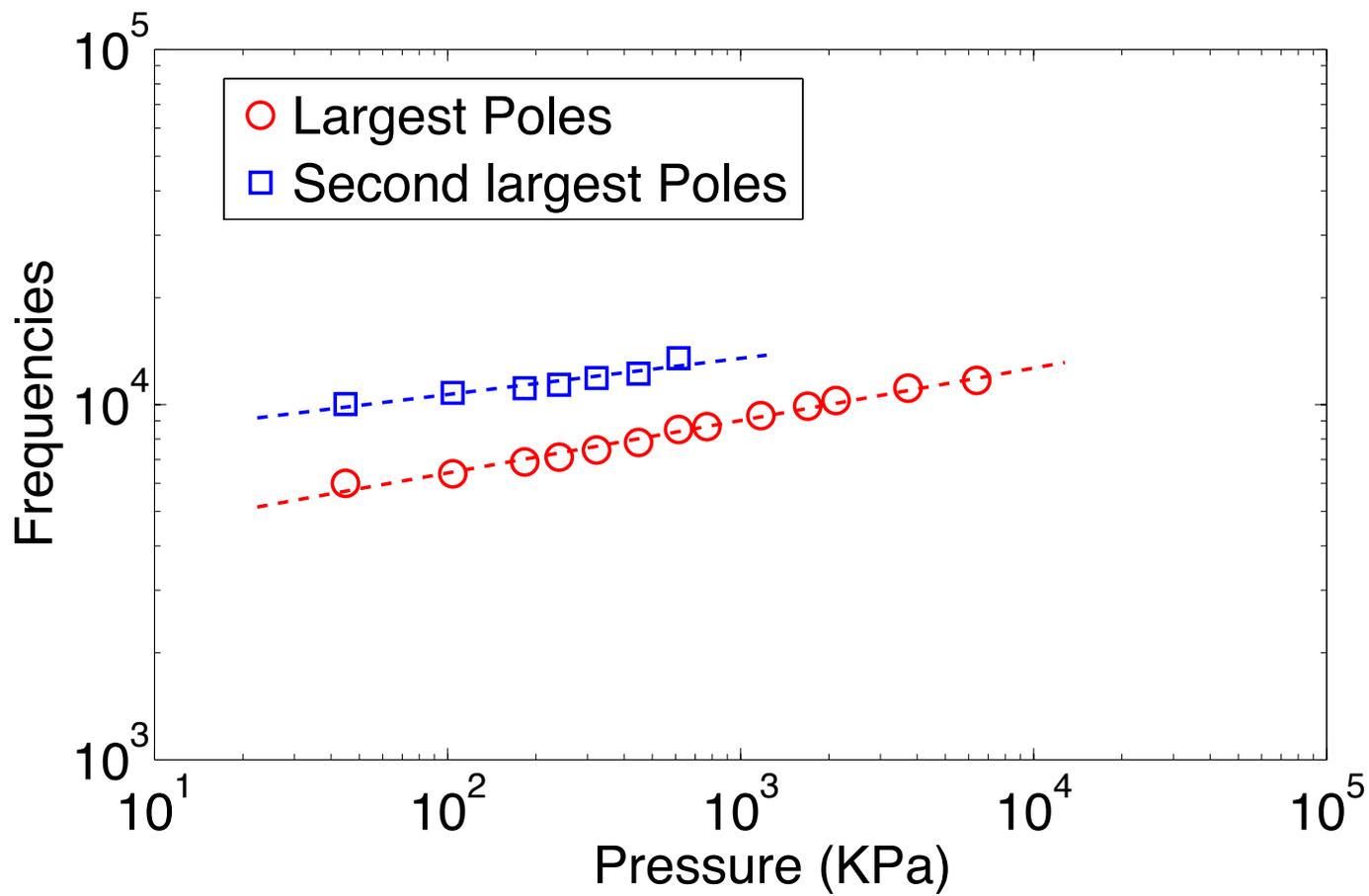}}
\caption{(Color online) Scaling of the modes obtained from the first and second peak
  of the tungsten system showing similar scaling with pressure.}
\label{othermodes}
\end{figure}

\begin{figure}
\centering{  \includegraphics[width=0.9\columnwidth]{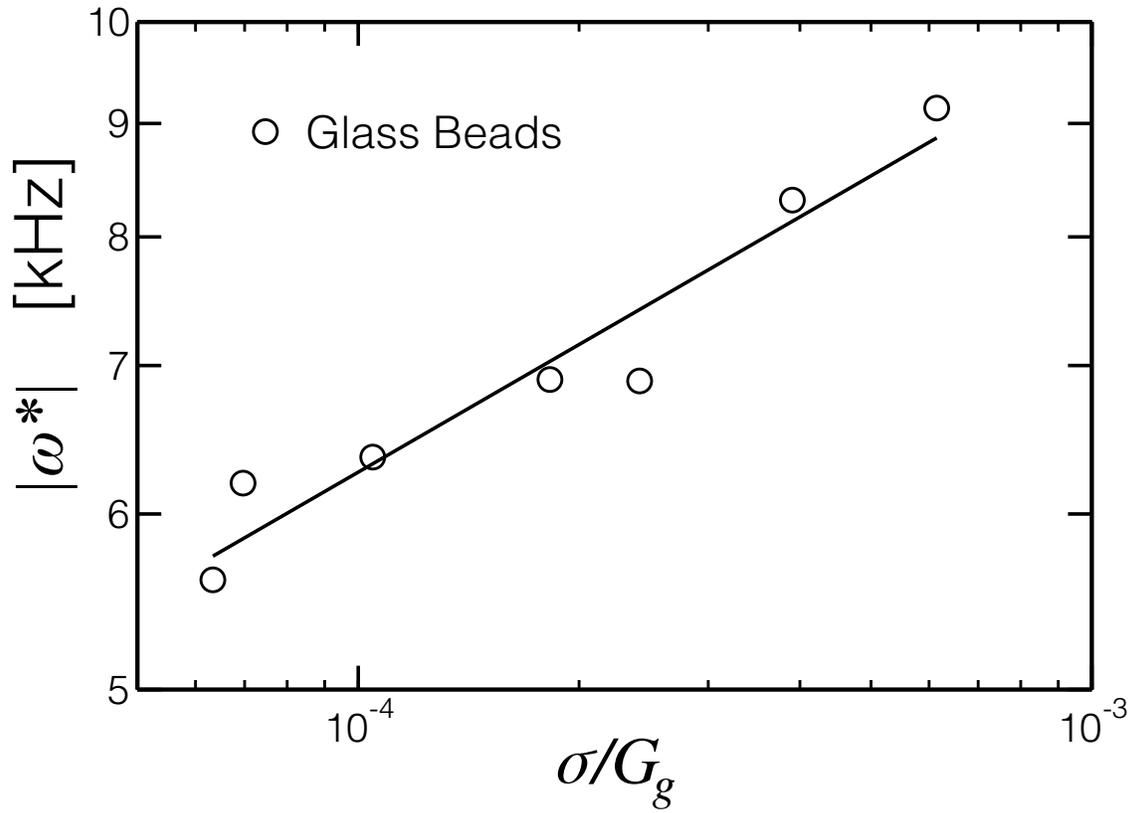}}
\caption{Test of reversibility in the glass beads experimental
  system. We follow a decreasing path in pressure and obtain similar
  exponent $\Omega'=0.20\pm0.02$ as the one obtained with the upward
  pressure path reported in Table \ref{table2}.}
\label{up-down}
\end{figure}

We also test the scaling of the other modes obtained from the
effective mass by looking at the modes with smaller residue than the
main characteristic frequency.
We find that the other modes scale with similar exponent as the
characteristic mode $|\omega^*|$ (Fig. \ref{othermodes}).  
The lack of data in the second peak for large stress is due
to the fact that they fall outside of our experimental range of
observation for high enough pressures. 
This is
also corroborated by the data collapse shown in Fig. \ref{experiment}c
and \ref{experiment}d.  The result of the characteristic frequency
following a decreasing path in stress is plotted in Fig. \ref{up-down}
showing that the scaling of the characteristic frequency with stress
yields similar exponent as the upward path in stress shown in
Fig. \ref{scaling}.





\section{Conclusions}

Our results suggest that granular materials near jamming behave
critically with a characteristic frequency defined by a critical
exponent $\delta$.  The experimental exponents $\delta$ are
consistently smaller than $\delta_{\rm th} = 1/2$ predicted by theory
\cite{Matthieu2005,Wyart2005}.  Such an anomalous exponent
$\delta<1/2$ poses new theoretical challenges to explain frictional
systems with translational and rotational degrees of freedom.  It
might be possible that the path followed in stress space may change
the value of the exponents. The usual path followed in previous works
is along the line of zero shear stress, as a function of packing
fraction. The present experiments follow a line of changing uniaxial
stress, which would place it along a path that excites bulk and shear
modes. Thus, it is plausible that the exponents might be affected by
the path taken to approach the jamming transition.  Another
interpretation is that the lowest of our frequencies may not be a
proxy for the cutoff frequency $\omega_D^*$, which separates the
region of excess density of states from the Debye region. However, if
some universality could be claimed at the jamming transition, both
characteristic frequencies should scale with stress with the same
universal exponent. In the same line, it is interested to note that
the lowest frequency estimated from the effective mass is directly
obtained from the dynamical matrix as done in $\omega_D^*$ as well.

In general, other mechanical properties such as elastic constants
(shear and bulk moduli), sound speeds, and attenuation can be obtained
from the effective mass measurement
\cite{Chaur2009,JValenza2012}. Furthermore, the theory can be extended to non-spherical particles as well \cite{Baule2}.Thus, the effective mass technique
facilitates a systematic test of the scaling laws of the anomalous
mechanical and acoustic properties of athermal and dissipative
granular systems near the jamming transition. We hope that this
technique may open new experimental tests of granular matter near
the jamming point.

\vspace{1cm}

{\bf Acknowledgements.} We acknowledge funding by DOE Office of Basic
Energy Sciences, Geosciences Division, Grant DE-FG02-03ER15458. This
work was partially supported by NSF-CMMT. We are grateful to S. Reis
and E. Kyeyune-Nyombi for help with experiments and simulations. F. Santibanez acknowledges FONDECYT POSTDOC project N$^o$ 3120130.


\begin{table*}[ht]
\centering
\begin{tabular}{c | c | c | c | c | c | c | c}
\hline Material & Size & Shape & Damping & $\Omega'$ & $\alpha$ &
$\Omega = \alpha \Omega'$ & $\delta$  \\ 
\hline Tungsten powder & $\sim 150\mu$m &
irregular & coating & 0.15 & 1.15 & 0.17 & 0.10  \\ 
Glass beads & 1 mm & sphere & no coating & 0.21 & 1.25 & 0.26 & 0.14\\ 
Steel balls & 2 mm & sphere & no coating & 0.195 & 1.21 & 0.24
& 0.13 \\ 
\hline Theory linear spring \cite{Wyart2005} & -- & sphere & -- & 1/2
& 1 & 1/2 & 1/2 \\ Theory Hertz \cite{Wyart2005} & -- & sphere & -- &
1/2 & 3/2 & 3/4 & 1/2 \\ 
\hline
\end{tabular}
\caption{\label{table2} Summary of the measured exponents and
  predicted by theory.}
\end{table*}

\clearpage

\section{Appendix. Experimental set up}
\label{setup}

We characterize the normal mode spectrum of the granular medium by
measuring the effective mass. The main results in the text are for a
granular medium made of tungsten powder that is lightly coated with
silicone oil (PDMS) of viscosity 894 cP.  We also present results for
spherical glass beads of 1 mm diameter and spherical steel balls of
2mm. Both systems are uncoated. The preparation protocol is the same
for all cases. Below, we explain the experimental procedure for
tungsten particles.

The tungsten powder \cite{JValenza2009} and coating procedure
\cite{JValenza2012} are discussed elsewhere. In the latter reference
we found that the light PDMS coating significantly dampened or
eliminated the low amplitude modes in the system. This is advantageous
for the purposes of this study because it permits us to unambiguously
monitor the effect of pressure on the trajectory of the dominant large
amplitude modes in the system. The purpose of using tungsten as a
material for the particles is to achieve a large mass of the granular
system in comparison with the mass of the cup, to maximize the
difference between both quantities and obtain a reliable effective
mass.

\begin{figure}
\centering{
   \includegraphics[width=0.9\columnwidth]{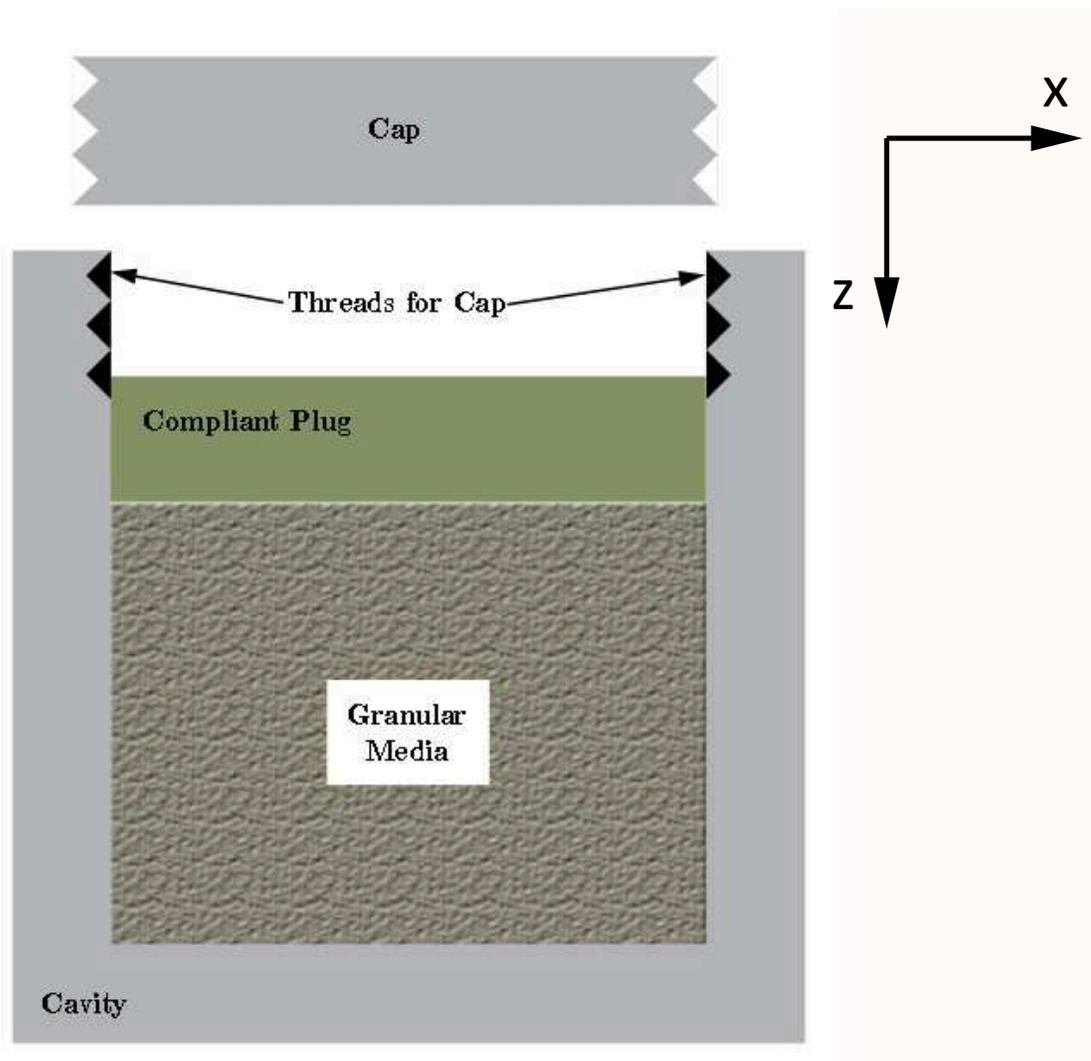} }
\caption{(Color online) Schematic of cup used in effective mass experiments. A
  compliant plug is sandwiched between the tungsten powder and the
  cap. The uniaxial confining stress is increased by tightening the
  cap.} \label{cappedCup}
\end{figure}

After coating, $\sim$100 g of the weakly wet powder is poured into a
cylindrical cup (Fig. \ref{cappedCup}). The cup has internal
diameter=2.54 cm, and height=3.08 cm. The sides of the cup are tapped
to encourage the powder to settle, and yield a flat free
surface. Finally, the system is compacted using a cyclical stress
imposed with an Inkstrom press of sequentially increasing then
decreasing amplitude as we have previously developed
\cite{brujic}. The maximum stress amplitude is 118.5 kPa. It was
previously shown \cite{JValenza2009} that this handling procedure
suitably produces a finite sized granular medium with a reproducible
effective mass which is independent on the amplitude of
oscillation. The mechanical compaction protocol is discussed in detail
in reference \cite{JValenza2009}.

The effective mass measurement is discussed in reference
\cite{Chaur2009,JValenza2009,JValenza2012}. The cylindrical cup is
subjected to a vertical, sinusoidal displacement where the vibrational
frequency is varied in the range 0.1 - 15 kHz. We take frequency steps
of 10 Hz, and at each frequency we measure the force on the bottom of
the cup, and the acceleration at two points on opposite ends of a
single diameter. The mass is determined by the difference between the
ratio of the force to the average acceleration, and the static mass of
the cup. The effective mass is a frequency dependent complex value,
owing to the partial in phase partial out of phase motion of the
individual grains.

The objective of this work is to observe the effect of a uniaxial
confining pressure on the dominant normal modes in our granular
system. To impose a sequentially increasing stress on the granular
medium, we place a thin $\sim$ 5 - 8 mm urethane plug on the free
surface of the powder, and screw a cap on top of the cup. After
determining the orientation of the cap corresponding to initial
contact with the top of the urethane, we further tighten the cap to
impose a uniaxial compressive stress. Each rotation of the cap
corresponds to an axial displacement that is governed by the
characteristics of the threads which support the cap. To convert this
displacement to a stress, we use a mechanical testing machine to
characterize the stress-displacement relationship of the coupled
urethane/tungsten powder system in the cup.

\begin{figure}
\centering{\includegraphics[width=1\columnwidth]{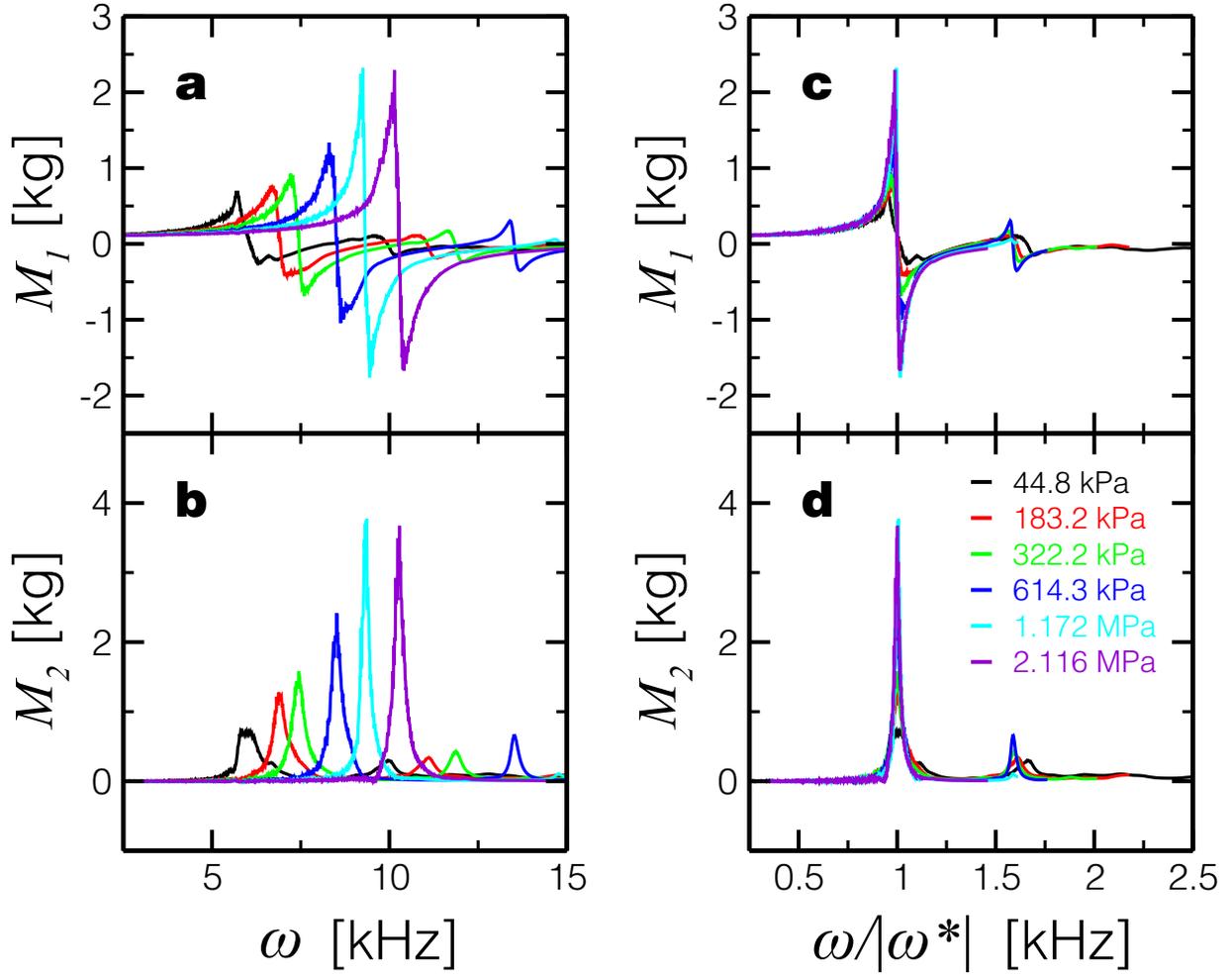}}
\caption{(Color online) Effective mass of tungsten particles in a cup as a
  function of driving frequency: {\bf (a)} Real part $M_1(\omega)$,
  and {\bf (b)} Imaginary part, $M_2(\omega)$, plotted for the
  indicated external stresses.  We calculate $\tilde{M}$ for 13
  stresses between $\sigma=44.8$ kPa and 6.39 MPa and show six curves
  as indicated for clarity. All datasets are available at
  http://jamlab.org. {\bf (c)-(d)} Data collapse of $M_1(\omega)$ and
  $M_2(\omega)$ according to the scaling scenario
  Eq. (\ref{critical}).  }
\label{experiment}
\end{figure}

\begin{figure}
\centering{ \includegraphics[width=0.9\columnwidth]{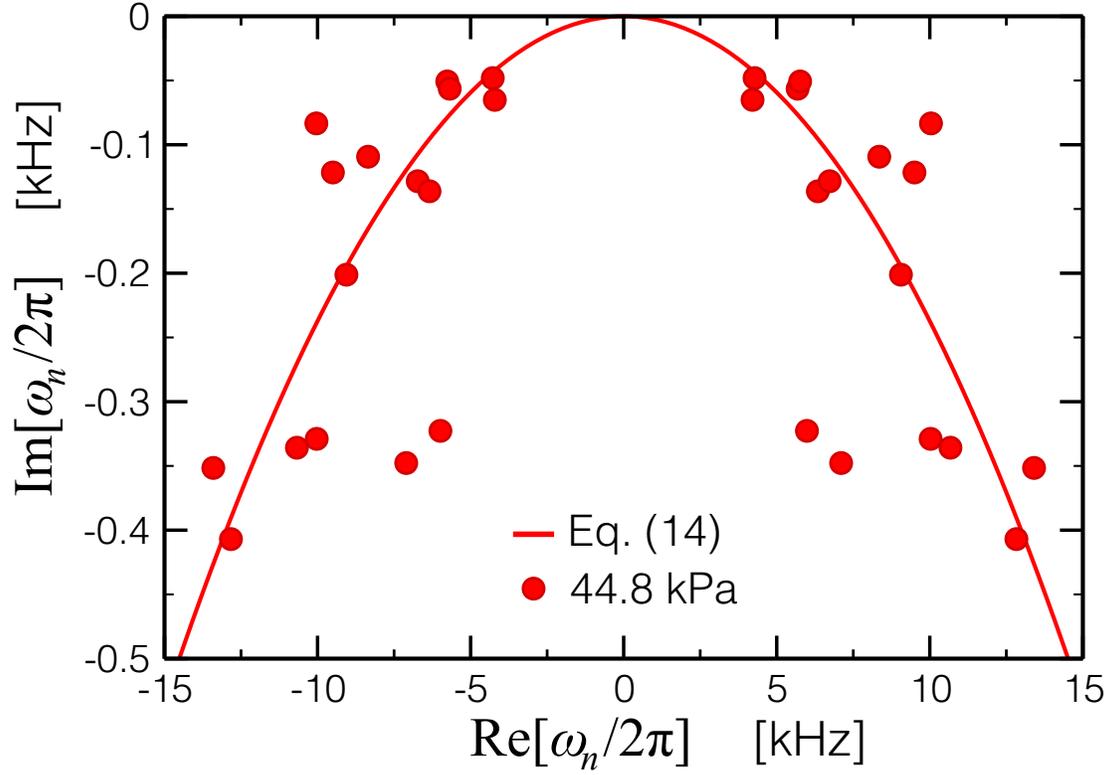}}
\caption{(Color online) Location of the normal mode frequencies in the
complex-value plane (Re[$\omega_n$], Im[$\omega_n$]) which contribute
to the pole decomposition Eq. (\ref{pd}) of the data plotted in
Fig. \ref{experiment} for $\sigma=$44.8 kPa. Solid line represents a
fit according to the parabolic Eq. (\ref{quadratic}). This equation
neglects the fluctuations in the force distribution, which explains
the scattering of the data around the parabolic fit.
Other stresses behave similarly.}
\label{residues}
\end{figure}

\begin{figure}
\resizebox{8.5cm}{!} {\includegraphics{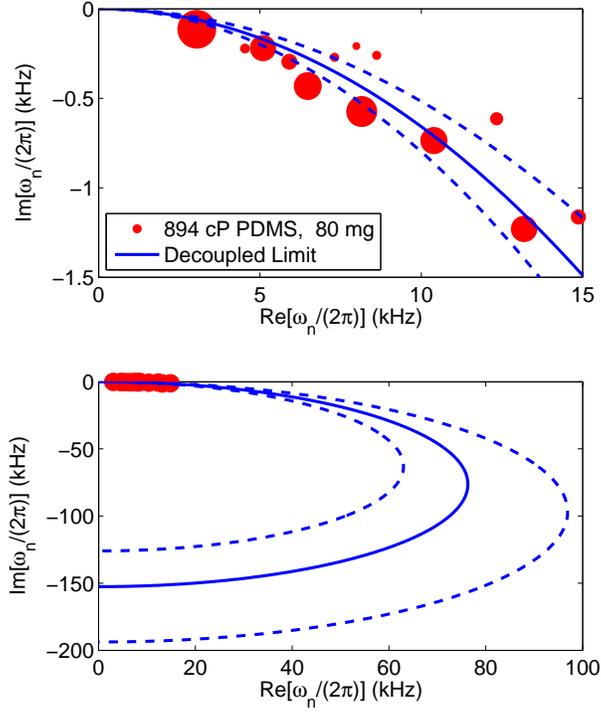}} \caption{(Color online) (a)
  Complex-valued normal mode frequencies determined from experimental
  data of loose granular tungsten particles, lightly coated with 80 mg
  of PDMS of viscosity 894 cP.  From Ref \cite{JValenza2012}. This
  system is identical to the system of Fig. \ref{residues} except for
  larger amount of PDMS which increases the dissipation. The size of
  each symbol is proportional to the magnitude of the residue, $A_n$,
  in the decomposition of the effective mass data.  The solid curve is
  a best fit of the data to Eq. (\ref{circle_plot}).  The dashed
  curves are for radius values corresponding to the full width at
  double the minimum of the cost function, Eq. (\ref{cost_function}).
  (b) Same as (a) but with an expanded scale.  We expect the missing
  normal mode frequencies - those whose residues are negligibly small
  in the measured effective mass - to lie within these bands or on the
  negative imaginary axis.} \label{New_fig}
\end{figure}

The results of the effective mass are shown in Fig.  \ref{experiment}.
It is visually apparent from Fig. \ref{experiment} that only a few
modes appear in the effective mass while 12N modes are expected. The
effective mass is sensitive to the extended modes in the system
corresponding to collective motion of the grains. For instance, the
main mode from where we obtain the characteristic frequency, is an
extended mode spanning the system size. The remaining modes are more
localized, correspond to modes with smaller residues, or are heavily
damped and are difficult to observe them in the shape of the effective
mass. However, all the 12N modes appear in the effective mass
measurement.
The remaining modes, being more damped, do not show easily in the
shape of the effective mass. Thus, the effective mass is most
effective in finding the main extended modes in the system.

\begin{figure}
\centering{
    \includegraphics[width=0.9\columnwidth]{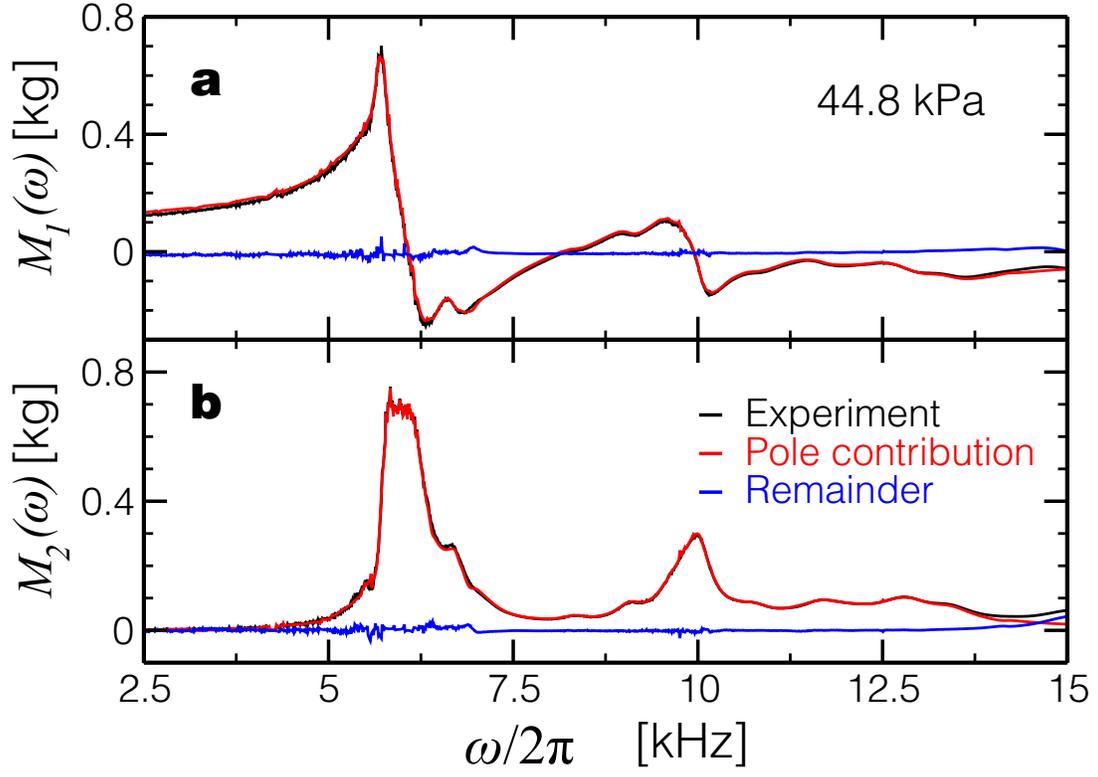}}
\caption{(Color online) Pole decomposition. Comparison of pole decomposition using
  Eq. (\ref{pd}) and the obtained poles and experimental data at 44.8
  kPa for the tungsten particles. We also plot the difference showing
  that many more nodes with small residues are still present in the
  effective mass.}
\label{fitting}
\end{figure}

Thus, modes with small residues are difficult to see in the effective
mass since they are overshadowed by the extended modes with larger
residues appearing as large peaks in Im($\omega$). This picture is
corroborated in the pole analysis of Fig. \ref{fitting}.  When we
subtract the main pole contribution of the characteristic frequency
$|\omega^*|$, a finer structure with many small peaks (called
Remainder in Fig. \ref{fitting}) appears. This finer structure
corresponds to the remainder modes with smaller residues. By
subtracting one by one the contribution of these modes from the
effective mass, in principle, we could obtain the entire subset of the
12N modes that appear in the window of observation. In practice, we do
it for only the few largest modes since we are interested in the
extended modes that test the criticality of the system. However, the
procedure can be extended to find more modes until a given resolution
preset by the experimental measurement of the effective mass. That is,
one can continue extracting modes from the pole decomposition as long
as the mass measurement can resolve the peaks in the Reminder of the
effective mass.

Thus, while nonlinearities are inherent in the system, still the 12N
modes are part of the effective mass, at least those that are within
the experimental window of observation.  As long as the sought-after
poles lie within the rectangle [$\pm 15$ kHz, $\pm i$15 kHz], our
procedure will determine their properties with a reasonable
accuracy.

\clearpage

\clearpage

\clearpage

\clearpage

\clearpage

\clearpage

\clearpage

\clearpage

\clearpage

\clearpage

\end{document}